\documentclass[prl,twocolumn,showpacs,nofootinbibs,superscriptaddress
]{revtex4}
\usepackage{epsf}
\usepackage{epsfig}
\usepackage{amsmath}

\newcommand{\be}{\begin{eqnarray}}
\newcommand{\ee}{\end{eqnarray}}
\newcommand{\ba}{\begin{array}}
\newcommand{\ea}{\end{array}}

\begin{document}
\preprint{}

\title{Chiral Inflation of the Pion Radius}

\author{Irina A. Perevalova}
 \affiliation{Physics Department, Irkutsk
State University, Karl Marx str. 1, 664003, Irkutsk, Russia }
\author{Maxim V. Polyakov}
\email{maxim.polyakov@tp2.rub.de}
\affiliation{Petersburg Nuclear Physics Institute,
Gatchina, St.\ Petersburg 188300, Russia} \affiliation{Institut
f\"ur Theoretische Physik II, Ruhr--Universit\"at Bochum, D--44780
Bochum, Germany}
\author{Alexander N. Vall}
 \affiliation{Physics Department, Irkutsk
State University, Karl Marx str. 1, 664003, Irkutsk, Russia }

\author{Alexei A. Vladimirov}
%\email{}
\affiliation{Institut f\"ur Theoretische Physik II,
Ruhr--Universit\"at Bochum, D--44780 Bochum, Germany}

\date{\today}

\begin{abstract}
We derive expression  for the large $b_\perp$ asymptotic of the 3D parton distributions $q(x,b_\perp)$ in the pion. The  asymptotic depends exclusively on the mass scales $F_\pi$ and $m_\pi$.
Therefore it provides us with a nice example of a
strict non-perturbative result for the partonic structure of Nambu-Goldstone bosons in QCD.
Analyzing the $x$-dependent pion transverse radius we reveal a new phenomenon of
``chiral inflation"-- in the parametrically wide region of Bjorken $x$ ($m_\pi \ll 4\pi F_\pi \sqrt x \ll 4\pi F_\pi$)
the pion radius grows exponentially fast with the rapidity $\eta=\ln\left(1/x\right)$. We show that the
partons in this interval of Bjorken $x$ contribute to famous logarithmic divergency of the pion radius. In other words, 
the partonic picture of the classical result of $\chi$PT is provided.
The phenomenon of the chiral inflation is at
variance with the Gribov diffusion, because of long-range interaction of the Nambu-Goldstone bosons.

\end{abstract}

\pacs{}

\maketitle
\thispagestyle{empty}

Famous experiments of R. Hofstadter et al. \cite{Hofstadter} on elastic electron scattering established
that hadrons are not point-like
particles and have a non-trivial spatial structure of a finite
size of the order of $\sim$1~fm.
For studies of the space-time picture of hadrons and their interactions in a quantum field theory
it is useful to consider the hadron in the infinite momentum frame \cite{gribov}. In this frame the hadron moves
with almost speed of light and therefore one can easily separate the partons (quark, anti-quarks and gluons)
which ``belong" to the hadron from that ``belonging" to vacuum fluctuations.

Quantitatively the partonic space-time structure of the hadron is described by so-called 3D parton distributions \cite{burk}, which can be obtained as the zero longitudinal
momentum transfer limit of generalized parton distributions (GPDs) (GPDs were introduced in \cite{pioneers},
for review and definitions see \cite{GPDrev}):

\be
q(x,b_\perp)=\int \frac{d^2\Delta_\perp}{(2\pi)^2}\ e^{-i \vec b_\perp \vec \Delta_\perp} H(x,\xi=0,-\Delta_\perp^2).
\label{limit}
\ee
Here $x>0$ corresponds to the quark distribution and $x<0$ to anti-quark distribution with minus sign in front.
The 3D parton distribution ($q(x,b_\perp)$) is of fundamental importance for understanding quark and gluon structure of hadrons. It gives the probability density to
find a parton (quark, anti-quark or gluon) with longitudinal momentum fraction $x$
and the coordinate $b_\perp$ in the transverse plane, in this way providing us with the three dimensional picture of a hadron.
The partons (to be specific we shall discuss below the distributions of up quarks and anti-quarks in $\pi^+$)
 with the longitudinal momentum fraction $x$ within the hadron occupy
a disc of the average transverse size squared given by:

\be
\label{bb}
b_\perp^2(x)=\int d^2 b_\perp\ b_\perp^2\ q(x,b_\perp)\, .
\ee
The C-odd transverse size of the hadron (for resent discussion see \cite{miller,miller2,vw}) which is determined by the
slope of the vector form factor at low momentum transfer can be
obtained by integrating $b_\perp^2(x)$ over the momentum fraction:
\be
\label{btot}
b_\perp^2=\int_{-1}^1 dx\ b_\perp^2(x)\, .
\ee

One can also introduce the normalized quark probability density in the transverse plane:
\be
\label{ndensity}
\rho(x,b_\perp)=\frac{q(x,b_\perp)}{q(x)}\,
\ee
where $q(x)=\int d^2 b_\perp q(x,b_\perp)$ is the forward quark distribution function.
Eq.~(\ref{ndensity}) defines the conditional probability density
to find quark around
the transverse distance $b_\perp$ if the longitudinal momentum fraction of the quark is fixed to $x$.
One can show on general grounds \cite{burk} that $\rho(x,b_\perp)\to \delta(\vec b_\perp)$ at $x\to 1$.

One can interpret
the normalized quark density (\ref{ndensity}) as an evolution of the probability density for a stochastic
motion of a particle in the transverse plane. The role of the evolution time is played by the rapidity $\eta=\ln(1/x)$.
At the initial time $\eta=0$ ($x=1$) the particle is localized at $\vec b_\perp=0$. For the stochastic process
we can introduce the mean square distance of the particle as follows:

\be
\label{dperp}
d_\perp^2(x)=\int d^2 b_\perp\ b_\perp^2\ \rho(x,b_\perp)=\frac{b_\perp^2(x)}{q(x)}.
\ee

V.N.~Gribov in his famous lectures \cite{gribov} derived that in a broad class of quantum field theories the stochastic process
discussed above corresponds to a Gaussian random walk in the transverse plane, hence one deals in this case with
usual diffusion:

\be
\label{diffusion}
d_\perp^2(x)=D\ \ln\left(\frac 1x \right)=D\ \eta,
\ee
where $D$ is a diffusion coefficient.
Gribov diffusion is realized in a simple (by Regge picture inspired) model for the 3D quark distributions \cite{GPDrev}.
\be
q(x,b_\perp)=\int \frac{d^2\Delta_\perp}{(2\pi)^2}\ e^{-i \vec b_\perp \vec \Delta_\perp}\ x^{\alpha^\prime \Delta_\perp^2}\ q(x),
\ee
where $\alpha^\prime$ is a slope of the corresponding Regge trajectory. It is obvious that in this model $d_\perp^2(x)$ obeys
the diffusion law (\ref{diffusion}) with the diffusion coefficient $D=4 \alpha^\prime$

Gribov diffusion (\ref{diffusion}) has been obtained assuming that the interaction in an underlying
field theory is {\it short range} and {\it strong}, see discussion in Ref.~\cite{gribov}.
These assumptions are not satisfied for the interaction of Nambu-Goldstone bosons.
Our aim here is to investigate the quark 3D distributions of the pion, the
Nambu-Goldstone boson of the spontaneous breakdown of the chiral symmetry in QCD.
Especially we are interested in the behaviour with $\eta=\ln(1/x)$ of the pion transverse radius.

The very fact of the spontaneous breakdown of the chiral symmetry in
QCD allows one to obtain a number of strict results on the large
distance (low-energy) behaviour of various quantities describing
hadrons and their strong interactions (for a historical review see
Ref.~\cite{Weinberg}).  Till recently the
applications of Chiral Perturbation Theory
($\chi$PT) have been devoted mostly to studies of soft interaction of hadrons and to
calculations of hadrons static properties (the status of $\chi$PT is reviewed in Refs.~\cite{reviews}).

One of classical results of $\chi$PT is the divergency of the pion radius in the chiral limit ($m_\pi\to 0$).
The corresponding radius behaves with the parametrically small pion mass as follows \cite{piradius}:

\be
\label{pir}
b_\perp^2=\frac{2}{3 \Lambda_\chi^2}\ \ln\left(\frac{\Lambda_\chi^2}{m_\pi^2}\right)\left[1+O\left(\frac{1}{\ln\left(\frac{\Lambda_\chi^2}{m_\pi^2}\right)}\right)\right]\, .
\ee
Here
\be\label{lambda}\Lambda_\chi=4\pi F_\pi \approx 6~{\rm fm}^{-1}\, \ee
 is the inherent
for $\chi$PT short distance scale ($1/\Lambda_\chi\approx 0.17$~fm).
In connection with the text book result (\ref{pir}) one can pose several questions.
Which range of Bjorken $x$  of quarks and antiquarks is responsible for the logarithmic divergency of the pion radius? What is the ``evolution" of the pion transverse radius
with increasing of the rapidity $\eta=\ln\left(1/x\right)$? Is the Gribov diffusion (\ref{diffusion}) still valid for the partons in the Nambu-Goldstone bosons? 

Various aspects of the $\chi$PT for parton distributions have been developed in
Refs.~\cite{sav,kiv02,che,Kivel:2004bb,man,Ando,gpd1,gpd2,Strikman:2009bd}.
Applications of the $\chi$PT to the partonic structure of hadrons revealed a number of new interesting phenomena.
In particular, in Refs.~\cite{gpd1,gpd2} it was demonstrated that
the standard $\chi$PT
should be modified when it is applied to (generalized) parton distributions.
It was shown that the $\chi$PT for GPDs possesses nontrivial expansion parameter
$\sim p^2 \ln(p^2)/x$ ($p^2$ stays for external soft momenta and/or $m_\pi^2$, $x$ is the Bjorken scaling variable).
Presences of such parameter makes the all-loop resummation of $\chi$PT for GPDs imperative, because for $x\sim p^2$ the new parameter is
not anymore small and we are confronted with problem of summation of large infrared logs. Indeed, for $x\sim O(p^2)$ the $n$th order contribution
to chiral expansion of GPDs
is parametrically proportional to $\left[p^2 \ln(p^2)/x\right]^n\sim \left[\ln(p^2)\right]^n$ and hence all orders of chiral expansion must be taken into account.
In Refs.~\cite{gpd1,gpd2} the origin of the contributions $\sim \left[p^2 \ln(p^2)/x\right]^n$ was identified and the way to sum up such contributions is demonstrated. The summation of such singular contributions for C-odd ($q-\bar q$) 3D quark distribution can be performed following Refs.~\cite{gpd1,gpd2} and it has the following form \footnote{The detailed
account of the calculations  will be given elsewhere soon.}: 

\begin{widetext}
\be
\label{main2}
q(x,b_\perp)&=&\frac{1}{\pi b_\perp^2\ A \ln(b_\perp^2\Lambda_\chi^2)}\
\int_{-1}^1d\alpha \frac{1}{\sqrt{1-\alpha^2}}
\int_0^1 \frac{d\beta}{\beta} Q(\beta)\\
\nonumber
&\times&
\left[
\sqrt{\frac {x}{\beta} \frac{b_\perp^2 \Lambda_\chi^2}{A\ln(b_\perp^2 \Lambda_\chi^2)}+m_\pi^2 b_\perp^2}
\ K_1\left(2\sqrt{\frac{\frac {x}{\beta} \frac{b_\perp^2 \Lambda_\chi^2}{A\ln(b_\perp^2 \Lambda_\chi^2)}+m_\pi^2 b_\perp^2}{1-\alpha^2}}\right)
\right]
\cdot
\left(1+O\left(\frac{1}{\ln(\Lambda_\chi^2 b_\perp^2)}\right)\right)\, .
\ee
Here $K_1(z)$ is the modified Bessel function, and the functions $Q(\beta)$ is expressed in terms of the forward parton distributions $ q(x)$ as follows:

\be
\label{Qlarge}
Q(\beta)&=&-\left(1-\frac B2\right)\ \beta \frac{d}{d\beta}     q{}(\beta)
+B\ \left[      q{}(\beta)-
\int_\beta^1d z      q{}\left(\frac{\beta}{z}\right)\right].
\ee
\end{widetext}
The constants $A$ and $B$ in Eqs.~(\ref{main2},\ref{Qlarge}) are related to the large order asymptotics
of leading chiral logarithms  (LL) for the {\it massless} $\pi\pi$ scattering (see for details Ref.~\cite{gpd2}). 
The LL coefficients  can be computed  practically to the
unlimited loop order
with help of recursive equation derived in Refs.~\cite{Kivel:2008mf,Kivel:2009az,Koschinski:2010mr,Polyakov:2010pt}.
Unfortunately the analytical solution of this recursive equation is not yet found. However, it can be solved in the limit of large number of Nambu-Goldstone bosons
($N\to\infty$), see Refs.~\cite{Kivel:2008mf,Kivel:2009az,Koschinski:2010mr,Polyakov:2010pt}. The corresponding limit gives the following result for the coefficients $A$ and $B$ in Eqs.~(\ref{main2},\ref{Qlarge}):
\be
A^{\rm Large\ N}=\frac{N}{2},\  B^{\rm Large\ N}=0. 
\ee
Numerical solution of the recursive equation of Refs.~\cite{Kivel:2008mf,Kivel:2009az,Koschinski:2010mr,Polyakov:2010pt}
for the case of $N=3$ gives the following values of the parameters: 
\be
A=1.149, \ B=0.482.
\ee

Eq.~(\ref{main2}) provides the leading large $b_\perp$ asymptotic of 3D parton distribution
which depends exclusively on the mass scales $\Lambda_\chi$ and $m_\pi$. Therefore the corresponding asymptotic is model independent and it provides us with a nice example of a
strict non-perturbative result for the partonic structure of Nambu-Goldstone bosons in QCD.
One can easily check that the expression (\ref{main2}) satisfies the positivity properties of the 
parton densities and also it commutes with the DGLAP evolution kernel. The latter means that the
DGLAP evolution of the forward distribution $ q(x)$ which enters the r.h.s of Eq.~(\ref{main2})
leads to correct evolution of $q(x,b_\perp)$ on the l.h.s. of Eq.~(\ref{main2}). 

Eq.~(\ref{main2}) in the region of $1/\Lambda_\chi\ll b_\perp \ll 1/m_\pi$ and $x\ll \Lambda_\chi^2/b_\perp^2$ simplifies considerably
(we show result for the normalized quark density (\ref{ndensity})):

%\begin{widetext}
\be
\rho(x,b_\perp)&\approx& \frac{1}{2\pi b_\perp^2}\ \frac{\left(A\ \ln(b_\perp^2\Lambda_\chi^2)\right)^{\omega-1}}{(b_\perp^2\Lambda_\chi^2)^\omega}\\
\nonumber
&\times&
\frac{\sqrt\pi\ \Gamma(\omega+2)^3}{(\omega+2)\Gamma\left(\omega+\frac 52 \right)}
\left[1-\frac{B}{2}\frac{\omega-1}{\omega+1}\right]\, .
\ee
%\end{widetext}
Here $\omega$ is the slope of the forward quark distribution at small $x$, ${q}(x)\sim 1/x^\omega$.
We see that  the very form of the  $b_\perp$-dependence of the quark densities depends on
 the high energy behaviour of the forward parton densities. It is a non-trivial example of interlacement 
of the chiral and partonic degrees of freedom. 

Inspecting  Eq.~(\ref{main2}), we  see
that there are the following regions of Bjorken $x$ in which the
3D quark densities have qualitatively different behaviour at large
$b_\perp$:
\begin{itemize}
\item[(I)]
$x\lesssim \frac{m_\pi^2}{\Lambda_\chi^2}$,
in this region the behaviour of the 3D parton densities at large $b_\perp$ is
governed by the mass scale given by $m_\pi$.
\item[(II)]
$m_\pi \ll \sqrt x\Lambda_\chi\ll \Lambda_\chi$, in this region
the behaviour of the 3D parton densities at large $b_\perp$ is
governed by the new intermediate mass scale $\sqrt x \Lambda_\chi$
which is, however, much smaller than the typical hadronic scale.
We call this region of Bjorken $x$ as ``region of chiral
inflation". We shall see below that the pion
radius very rapidly (exponentially fast with increase of the
rapidity) inflates from typical hadronic size of $\sim
1/\Lambda_\chi$ to parametrically large size of the ``pion cloud"
$\sim 1/m_\pi$.
\end{itemize}

In the region (II) ($m_\pi \ll \sqrt x\Lambda_\chi\ll \Lambda_\chi$) we can neglect the pion mass in Eq.~(\ref{main2}).
Computing after that the $x$-dependent transverse radius (\ref{bb}) we obtain for the region (II) simple and elegant result:
\be
\label{binflation}
b_\perp^2(x)=\frac{2}{3\Lambda_\chi^2}\ \frac 1x \int_0^1d\beta\     {q}(\beta)= \frac{2}{3\Lambda_\chi^2}\ \frac 1x\, .
\ee
Here we use that the C-odd usual quark distribution is normalized
by $\int_0^1d\beta\     {q}(\beta)=1$. We see that in the region of $m_\pi \ll \sqrt x\Lambda_\chi\ll \Lambda_\chi$
(region (II))
the transverse pion radius inflates exponentially fast with the rapidity $\eta=\ln\left(1/x\right)$ ($b_\perp^2(\eta)\sim e^\eta$). 
We see that on the interval (II) the pion transverse radius grows from the typical hadronic radius of 
$b_\perp\sim 1/\Lambda_\chi$ to the size of the ``pion cloud" $b_\perp\sim 1/m_\pi$.  Integrating Eq.~(\ref{binflation}) over the $x$ interval (II)
we obtain the following contribution to the total transverse pion radius (\ref{btot}):
\be
b_\perp^2=\frac{2}{3\Lambda_\chi^2}\ \int_{m_\pi^2/\Lambda_\chi^2}^1 \frac{dx}{x}=\frac{2}{3 \Lambda_\chi^2}\ \ln\left(
\frac{\Lambda_\chi^2}{m_\pi^2}\right).
\ee
That is exactly the classical result (\ref{pir}) for the pion radius in $\chi$PT. We come to the conclusion that 
the log divergency of the pion radius in $\chi$PT comes from the partons with Bjorken $x$ in the 
interval of $m_\pi \ll \sqrt x\Lambda_\chi\ll \Lambda_\chi$. 

If we now compute the normalized radius $d_\perp$ (\ref{dperp}) in the ``inflation region" (II) we obtain 
the exponential behaviour for large rapidities:
\be
\label{dinflation}
d_\perp^2(\eta)\sim \frac{1}{\Lambda_\chi^2} e^{(1-\omega) \eta},
\ee
where $\omega$ is the slope of the C-odd quark distribution at small $x$, $    {q}(x)\sim 1/x^\omega$ ($\omega\sim
0.5$).
The result (\ref{dinflation}) is obviously valid also for massless pions.

We see that the behaviour of the normalized pion radius (\ref{dinflation}) is at variance with the Gribov
diffusion (\ref{diffusion}). The reason for this is the fact that we obtain the ``inflation behaviour" (\ref{dinflation}) 
in the region where the pion mass can be neglected. For the massless limit the derivation of the Gribov
diffusion \cite{gribov} does not work as the assumption of short range interaction is violated. We note
that the exponential inflation of the radius (\ref{dinflation}) is valid for all Bjorken $x$ if the Nambu-Goldstone
boson is exactly massless. This can lead to anomalously large interaction of {\it massless} Nambu-Goldstone
bosons at high energies. 

For the non-zero pion mass the chiral ``inflation" of the pion radius (\ref{dinflation}) stops at $x\sim m_\pi^2/\Lambda_\chi^2$
when the pion radius reaches the size of the ``pion cloud" of $\sim 1/m_\pi$.  From Eq.~(\ref{main2}) we can obtain
that for $x\ll m_\pi^2/\Lambda_\chi^2$ the normalized transverse pion size $d_\perp$ is frozen at the value of order:
\be
d_\perp^2(\eta)\sim \frac{1}{\Lambda_\chi^2} \left[\frac{m_\pi^2}{\Lambda_\chi^2}
\ln\left(\frac{\Lambda_\chi^2}{m_\pi^2}\right)\right]^{\omega-1}, \   {\rm for}\  \eta\gg \ln\left(\frac{\Lambda_\chi^2}{m_\pi^2}\right).
\ee
We encounter an interesting phenomenon -- the $m_\pi$ dependence of the saturation radius depends 
on the values of the slope  of the parton distribution $\omega$. As the result the $m_\pi^2$ dependence of the saturation
radius is non-analytical with the non-trivial exponent. It is nice example of how the chiral counting depends 
in non-trivial way on the high energy behaviour of the parton distributions. 
The contribution of the interval $x\lesssim \frac{m_\pi^2}{\Lambda_\chi^2}$ to the total transverse pion radius $b_\perp$ is of order $1/\Lambda_\chi$, i.e. this interval is not chirally 
enhanced in the limit of $m_\pi\to 0$.

To conclude, we derived expression (\ref{main2}) for the large $b_\perp$ asymptotic of the 3D parton distributions $q(x,b_\perp)$ in the pion. The corresponding asymptotic is model independent and it provides us with a nice example of a
strict non-perturbative result for the partonic structure of Nambu-Goldstone bosons in QCD.
Analyzing the $x$-dependent pion transverse radius we revealed a new phenomenon of
``chiral inflation"-- in the parametrically wide region of Bjorken $x$ ($m_\pi \ll \sqrt x\Lambda_\chi\ll \Lambda_\chi$)
the pion radius grows exponentially fast with the rapidity $\eta=\ln\left(1/x\right)$. We showed that the
partons in this interval contributes to famous log divergency of the pion radius (\ref{pir}). This divergency is a classical 
result of  $\chi$PT \cite{piradius}. Here we provided the partonic picture of
this classical result. We  revealed which partons are responsible for the divergency.
The phenomenon of the chiral inflation is at
variance with the Gribov diffusion, as the Nambu-Goldstone bosons experience long-range interactions.

\subsection*{Acknowledgements}
We are thankful to N.A. Kivel for many illuminating discussions and to Yu. L.~Dokshitzer for correspondence.
The work is supported in parts by German Ministry for Education and Research (grant 06BO9012) and by Russian Federal Programme ``Research and Teaching Experts in Innovative Russia"
(contract 02.740.11.5154).

\end{document}